\documentclass[reprint, amsmath, amssymb, aps, showpacs, showkeys, prl]{revtex4-1}

\usepackage{natbib}
\usepackage{graphicx}
\usepackage{dcolumn}
\usepackage{bm}
\usepackage{hyperref}
\usepackage{times}
\usepackage[english]{babel}

\ifpdf\usepackage{epstopdf}\fi

\def\oH{\omega_{\scriptscriptstyle \rm H}}
\def\op{\omega_{p}}
\def\oI{\omega_{\scriptscriptstyle \rm I}}
\def\oII{\omega_{\scriptscriptstyle \rm II}}
\def\orr{\omega_{\rm r}}
\def\o{\omega}

\begin{document}

\title{Multiple Scattering of Electromagnetic Waves by an Array of Parallel Gyrotropic Rods}
\author{V.~A.~Es'kin$^{1}$}
\email{vasiliy.eskin@gmail.com}
\author{A.~V.~Kudrin$^{1}$}
\email{kud@rf.unn.ru}
\author{T.~M.~Zaboronkova$^{2}$}
\author{C.~Krafft$^{3}$}
\affiliation{$^{1}$Department of Radiophysics, University of Nizhny Novgorod, 23 Gagarin Ave., Nizhny Novgorod 603950, Russia\\
$^{2}$Department of Nuclear Physics, Technical University of Nizhny Novgorod, 24 Minin St., Nizhny Novgorod 603950, Russia\\
$^{3}$Laboratoire de Physique des Plasmas, \'{E}cole Polytechnique, 91128 Palaiseau Cedex, France}

\begin{abstract}
We study multiple scattering of electromagnetic waves by an array of parallel gyrotropic circular rods and show that such an array can exhibit fairly unusual scattering properties and provide, under certain conditions, a giant enhancement of the scattered field.
Among the scattering patterns of such an array at its resonant frequencies, the most amazing is the distribution of the total field in the form of a perfect self-similar structure of chessboard type. The scattering characteristics of the array are found to be essentially determined by the resonant properties of its gyrotropic elements and cannot be realized for arrays of nongyrotropic rods. It is expected that the results obtained can lead to a wide variety of practical applications.
\end{abstract}


\maketitle
Multiple scattering of electromagnetic waves by periodically spaced elements demonstrates many intriguing features that are of considerable practical and scientific significance~\cite{Yasumoto2006,Joannopoulos2008,Wang2008,Liu2008,Chui2010}. Knowledge of the individual properties and mutual location of elements in a periodic structure illuminated by an incident electromagnetic wave makes it possible to determine the spatial distribution and other characteristics of the scattered field. At a fixed frequency, the properties of periodic structures can be controlled by a change in the positions and material parameters of the scattering elements. In many cases, it is almost impossible to change the internal geometry of a periodic structure such as a photonic crystal, for example, without its destruction or serious damage. However, the properties of the scattering elements in the array can easily be controlled if they consist of resonant gyrotropic materials, the parameters of which in some frequency ranges can be very sensitive to even slight variations in an external dc magnetic field. Such a possibility opens up new promising prospects for controlling the scattering characteristics of an array of such elements without changing their sizes or positions.
Despite significant progress in the analysis of multiple scattering by periodically spaced resonant nongyrotropic elements~\cite{Bever1992, Kurin2009}, the results obtained for them cannot be used for arrays of gyrotropic scatterers because of the fundamental difference between the scattering characteristics of nongyrotropic and gyrotropic elements. Although gyrotropic photonic crystals have recently been discussed in the literature~\cite{Wang2008,Liu2008}, most works on the subject do no consider the role of individual resonant properties of the periodically spaced gyrotropic elements constituting the array in the formation of the scattering pattern. Meanwhile, one can expect that the joint contribution of the individual and collective resonance scattering mechanisms to the diffracted field of arrays containing gyrotropic elements should lead to many interesting phenomena, the features of which are yet to be determined.  In this work, the scattering of a normally incident plane electromagnetic wave by a two-dimensional array consisting of parallel resonant gyrotropic circular rods is considered and the unusual scattering properties of such a structure are revealed and discussed.

\begin{figure}[h]
\includegraphics[height=33mm]{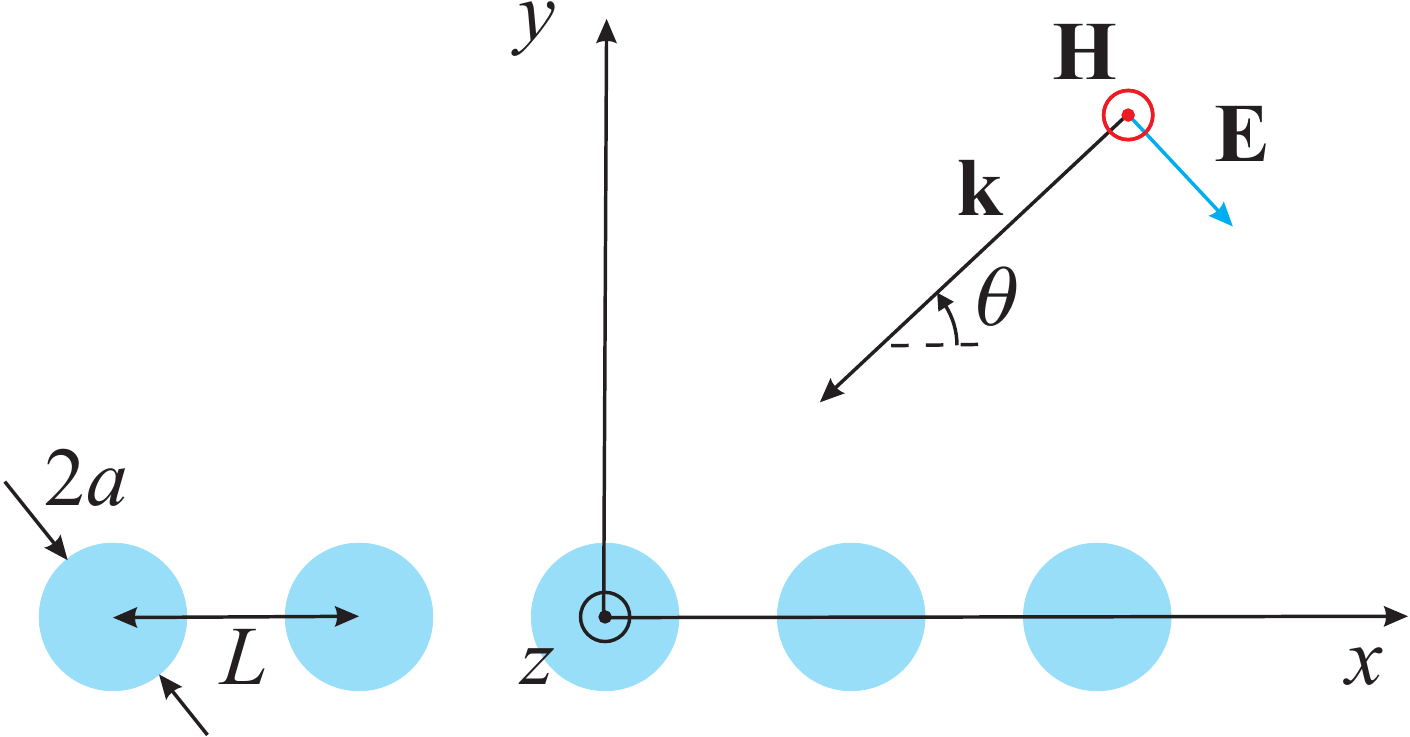}
\caption{\label{fig1}(color online) Geometry of problem.}
\end{figure}

Consider an equidistant array of identical parallel rods of radius $a$ (see Fig.~\ref{fig1}). The rods are embedded in a uniform background medium and aligned with an external magnetic field which is parallel to the $z$-axis of a Cartesian coordinate system ($x, y, z$). The axes of the rods lie in the $xz$ plane and are specified by the relations $x = j L$ and $y=0$, where $L > 2 a$ and $j=0,\pm 1, \pm 2,\ldots$~. The medium inside each rod is described by the permittivity tensor $\bf{\hat{\varepsilon}}$
which is typical of a magnetoplasma and has the following nonzero elements:
$\varepsilon_{\rho\rho}=\varepsilon_{\phi\phi}=\epsilon_{0}\varepsilon$,
$\varepsilon_{\rho\phi}=-\varepsilon_{\phi\rho}=-i\epsilon_{0}g$, and
$\varepsilon_{zz}=\epsilon_{0}\eta$.
Here, $\epsilon_0$ is the permittivity of free space,  $\varepsilon = 1 - \op^2/(\o^2 - \oH^2)$, $g= \op^2\oH/[(\o^2-\oH^2)\o]$, and $\eta=1-\op^2/\o^2$, where $\op$ and $\oH$ are the plasma frequency and the gyrofrequency of electrons, respectively, and $\o$ is the angular frequency. The  medium outside the rods is isotropic and has the dielectric permittivity $\varepsilon_{\rm out} = \epsilon_0 \tilde{\varepsilon}$. The incident wave is assumed to be a TE monochromatic plane wave whose magnetic field is polarized in the $z$-direction. We will not consider the incidence of a TM wave because its scattering is unaffected by the gyrotropic properties of the rods and similar to that in the case of isotropic rods. The wave vector $\bf{k}$ of the incident wave has the components $k_x = -k \cos\theta$, $k_y = -k \sin\theta$, and $k_z = 0$, where $k = k_0 \tilde{\varepsilon}^{1/2}$ is the wave number in the surrounding medium ($k_0$ is the wave number in free space) and $\theta$ is the angle of incidence, which is reckoned from the positive direction of the $x$-axis.

Omitting a time factor of $\exp(i\omega t)$, the total magnetic field normalized to the incident-wave amplitude can be sought in the form
\begin{equation}\label{eq1}
H_z = e^{-i\textbf{k}\textbf{r}} + \sum\limits_{j = -\infty}^{\infty}\sum\limits_{m = -\infty}^{\infty} D_{j,m} H^{(2)}_m (k \rho_j)e^{-i m \phi_j},
\end{equation}
where $m$ is the azimuthal index ($m = 0,\pm 1, \pm 2,...$), $D_{j,m}$ is the multipole coefficient characterizing the scattering by the $j$th rod to the $m$th azimuthal harmonic of the field, $H_m^{(2)}$ is the Hankel function of the second kind of order $m$, $
\rho_j$ is the distance from the axis of the $j$th rod to the observation point in the incidence plane, and $\phi_j = \arcsin(y/\rho_j)$. In Eq.~(\ref{eq1}), the first term represents the field of the incident wave, while the other terms account for the scattered field.

The magnetic field $H_{z;j,m}$ of the $m$th azimuthal harmonic inside the $j$th rod is written as
\begin{eqnarray}\label{eq2}
H_{z;j,m} = B_{j,m} J_m(k_0 q \rho_j),
\end{eqnarray}
where the multipole coefficient $B_{j,m}$ characterizes the field of the corresponding harmonic, $q = [(\varepsilon^2 - g^2)/\varepsilon]^{1/2}$, and $J_m$ is the Bessel function of the first kind of order $m$.

Satisfying the boundary conditions for the tangential field components $H_{z;j,m}$ and $E_{\phi;j,m}$ on the surface of the $j$th rod and using the standard technique based on the scattering matrix method~\cite{Yasumoto2006,Smigaj2009}, we can exclude the coefficients $B_{j,m}$ and obtain a system of equations for the scattering coefficients $D_{j,m}$ in the form
\begin{eqnarray}\label{eq3}
 &&S_m^{-1} D_{j,m}\!\! = \!\! \sum\limits_{n = -\infty}^{\infty} \Big[ \sum\limits_{l < j}^{} D_{l,-n} (-1)^{m+n} H^{(2)}_{m-n}(k L |j-l|) \nonumber\\
  &&+ \sum\limits_{l > j}^{} D_{l,-n} H^{(2)}_{m-n}(k L |j-l|) \Big] +  i^m e^{i ({m \theta} + k L j \cos\theta)},
\end{eqnarray}
where $S_m$ is the single-rod scattering coefficient:
\begin{eqnarray}\label{eq4}
&& S_m = -\frac{{J'_m}(\tilde{Q}){J}_{m}(Q) - {\tilde{\varepsilon}}^{1/2}{J}_{m}(\tilde{Q}){\cal{E}}_{m}}{{{H}^{(2)\prime}_m}(\tilde{Q}){J}_{m}(Q) - {\tilde{\varepsilon}}^{1/2}{H}_{m}^{(2)}(\tilde{Q}){\cal{E}}_{m}}.
\end{eqnarray}
Here,
$ {\cal{E}}_{m} = ({\varepsilon^2 - g^2})^{-1}\left[\varepsilon q J'_m(Q) + m {g}{(k_0 a)^{-1}}J_m(Q)\right],$
$ Q = k_0 q a$, $\tilde{Q} = k_0 \tilde{\varepsilon}^{1/2} a$, and the prime denotes the derivative with respect to the argument.

The translational symmetry of the problem makes it possible to use the discrete Fourier transform (with respect to $j$) and its inverse:
\begin{eqnarray}\label{eq5}
&& D_m(h)  = \sum\limits_{j={-\infty}}^{\infty} D_{j,m}e^{-i h L j },\nonumber\\
&& D_{j,m} =  \frac{L}{2\pi} \int_{-\pi/L}^{\pi/L} D_m(h)e^{ihLj} d h.
\end{eqnarray}

The Fourier transform of the incident-wave field comprises the Dirac function $\delta(hL - kL \cos{\theta})$. Therefore, we will seek $D_m(h)$ in the form $D_m(h) = \hat{D}_m \delta(hL - kL \cos{\theta})$. As a result, we arrive at the following system of equations for $\hat{D}_m$:
\begin{eqnarray}\label{eq6}
 S_{m}^{-1} \hat{D}_{m} = i^m e^{i m \theta} + \sum\limits_{n = -\infty}^{\infty} \hat{D}_{-n} G_{m-n},
\end{eqnarray}
where
\begin{eqnarray}\label{eq7}
\hspace{-5mm} G_{m} = \sum\limits_{l=1}^{\infty} H^{(2)}_{m}(k L l)[e^{i k L l \cos\theta} + (-1)^{m}e^{-i k L l \cos\theta}].
\end{eqnarray}

If the rods are electrically small such that $ka\ll 1$, we can restrict ourselves to the dipole approximation and retain only the terms with $m=\pm 1$ and $n =\pm 1$ in Eq.~(\ref{eq6}). Then from Eq.~(\ref{eq6}) we have
\begin{eqnarray}\label{eq8}
\hat{D}_{\pm 1} &=& \pm i \frac{[(S_{\mp 1}^{-1} - G_2) e^{\pm i\theta} - G_0 e^{\mp i \theta}]}{(S_{-1}^{-1} - G_2)(S_{1}^{-1} - G_2) - G_0^2}.
\end{eqnarray}

\begin{figure}[h]
\includegraphics[height=105mm]{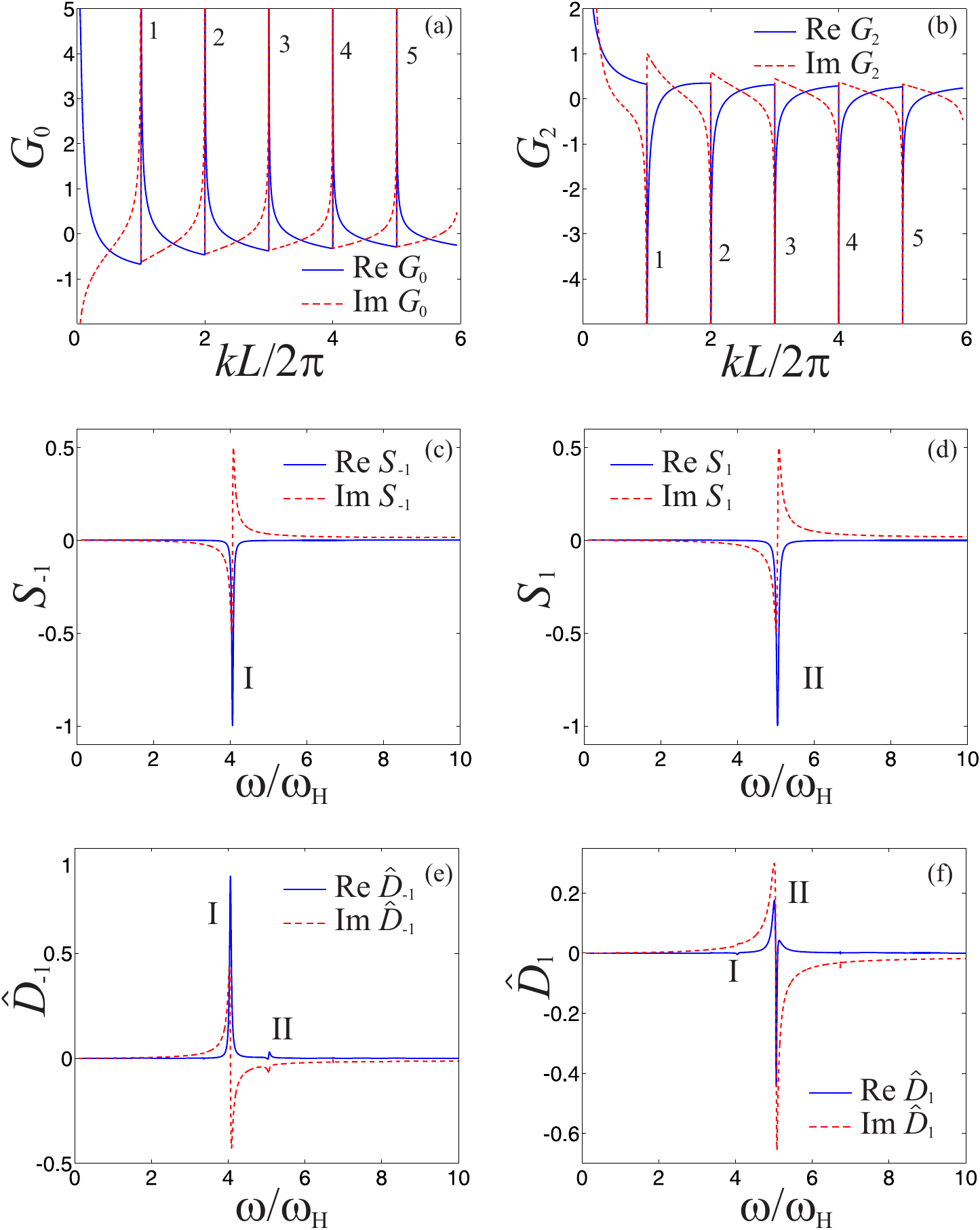}
\caption{\label{fig2}(color online) Frequency dependences of the real and imaginary parts (solid and dashed curves, respectively) of the array factors (a) $G_0$ and (b) $G_2$,  the single-rod scattering coefficients (c) $S_{-1}$ and (d) $S_1$, and the coefficients (e) $\hat{D}_{-1}$ and (f) $\hat{D}_{1}$.  The resonances of $G_{0,2}$ are labeled in order of increasing frequency. The resonances of $S_{-1}$ and $S_1$ at the frequencies $\oI$ and $\oII$ and the corresponding resonances of $\hat{D}_{-1}$ and $\hat{D}_{1}$ are denoted as $\rm I$ and $\rm II$, respectively. The values of parameters are chosen to be $\op/\oH = 6{.}47$, $\op a/c = 0{.}18$, $L/a = 132{.}8$, $\tilde{\varepsilon} = 1$, and $\theta = \pi/2$.}
\end{figure}

The array factors $G_0$ and $G_2$ in Eq.~(\ref{eq8}) have resonances in the cases $kL \cos\theta = 2\pi\nu$ for $|\theta| < \pi/2$ and $kL = 2\pi \nu$ for $|\theta| = \pi/2$, where $\nu = 1,2,\ldots$~. We will denote the corresponding resonant frequencies by $\o_{\nu}$. Figure~\ref{fig2} shows the frequency dependences of the quantities in Eq.~(\ref{eq8}) for $\theta = \pi/2$. It is seen in Figs.~\ref{fig2}(a) and \ref{fig2}(b) that $G_0$ and $G_2$ are discontinues functions of $kL$. When approaching the resonant value of $kL$ from the side of larger $kL$, the real and imaginary parts of $G_m$ tend to infinity and certain finite values, respectively. But if the resonant value of $kL$ is approached from the side of smaller $kL$, then the real and imaginary parts of $G_m$ tend to some finite values and infinity, respectively. Such behavior no longer takes place if the array is nonequidistant, the number of rods in the array is finite, or the surrounding medium is lossy.

In the case where $ka\ll 1$ and $k_0|q|a\ll 1$, the single-rod scattering coefficients $S_{-1}$ and $S_{1}$ have the resonant frequencies $\oI$ and $\oII$, respectively, which are located on different sides of the surface plasmon resonant frequency $\omega_{\rm r}$ of an isotropic plasma column~\cite{Crawford1963}. The frequencies $\oI$  and $\oII$ are approximately determined by the equation
$({\varepsilon +} g \chi_\mp)\left[1 + {(k_0 a)^2}/{2}\right]  = -\tilde{\varepsilon} $,
where $\chi_{\mp} = \pm 1$ for $m=\mp 1$.  It can be shown from this equation that under the additional condition $\tilde{\varepsilon}=1$, one obtains $\omega_{\rm r}=\op/\sqrt{2}$, $\op/2<\oI<\op/\sqrt{2}$, and $\oII > \op/\sqrt{2}$. With decreasing external magnetic field ($\oH \rightarrow 0$), the resonant frequencies $\oI$ and $\oII$ tend to $\op/\sqrt{2}$. With increasing external magnetic field (for $\oH \gg \op$), $\oI \rightarrow \op/2$ and $\oII \rightarrow \oH$.

Interaction of the individual and collective mechanisms of scattering can lead to both an increase and decrease in the scattering from the array compared with the scattering by a single rod. To demonstrate this fact, Figs.~\ref{fig2}(e) and \ref{fig2}(f) show the frequency dependences of $\hat{D}_{\pm 1}$ in the case where the frequency $\o_{3}$ of the $\nu = 3$ collective resonance of the array coincides with the resonant frequency $\oII$ of a single gyrotropic rod. In this case, the coefficient $\hat{D}_{-1}$ has the most pronounced resonance peak at a frequency that is very close to, but slightly lower than the frequency $\oI$, along with much less pronounced peaks at $\oII$ and $\omega_{\nu}$. At the same time, the coefficient $\hat{D}_{1}$ has minor peaks near $\oI$ and $\omega_{\nu}$. However, the behavior of $\hat{D}_{1}$ in the vicinity of $\oII$ differs significantly from that of $\hat{D}_{-1}$ near $\oI$.
An important feature of the coefficients $\hat{D}_{\pm 1}$ is that they decrease and become comparable, so that the scattering occurs as in the case of isotropic rods, if the frequency $\o$ tends to any of the quantities $\omega_{\nu}$ from the side of the higher frequencies. In this limit, the coefficients $\hat{D}_{\pm 1}$ turn out to be very small and the array becomes almost transparent for the incident radiation.

\begin{figure}[h]
\includegraphics[height=70mm]{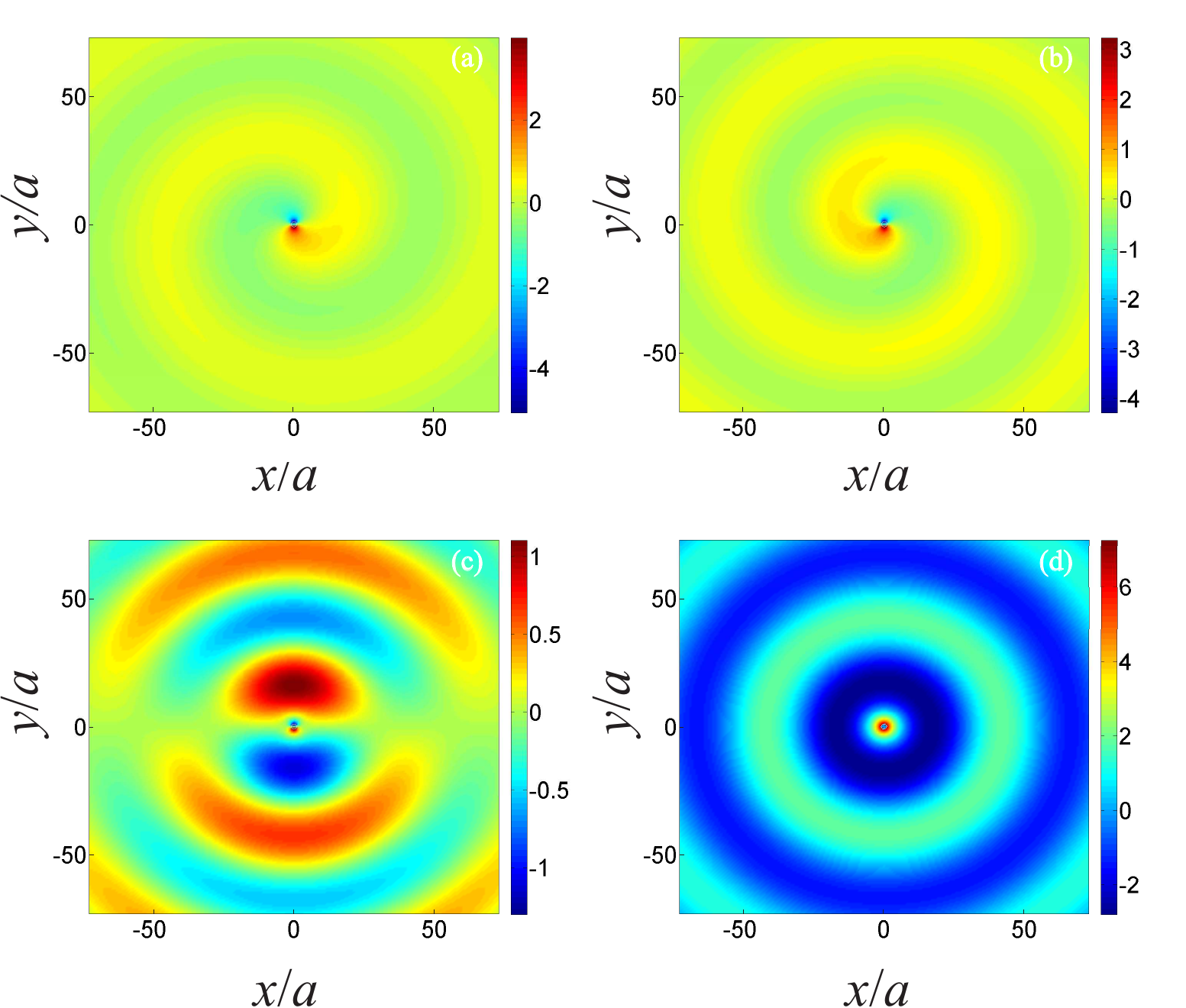}
\caption{\label{fig3}(color online) Snapshots of the scattered field $H_z^{\rm sc}$ in the cases where a plane waves is incident at an angle $\theta=\pi/2$ on a single gyrotropic rod ($j=0$) at the frequencies (a) $\o = \oI\simeq 4\oH$ and (b) $\o = \oII\simeq 5\oH$ and (c)~on an isotropic rod at the frequency $\o = \orr$. (d) The snapshot of the $m=0$ azimuthal harmonic of the field scattered by an isotropic rod at $\o = \orr$. The parameters $\op/\oH$, $\op a/c$, and $\tilde{\varepsilon}$ are the same as in Fig.~2.}
\end{figure}

Generally, the field scattered by an array of gyrotropic rods differs significantly from that in the case where an array consist of isotropic cylindrical scatterers. Before proceeding in comparison of these two cases, we turn to the fields scattered by a single rod when the incidence angle $\theta = \pi/2$. Figures~\ref{fig3}(a) and \ref{fig3}(b) show  the snapshots of the scattered-field component $H_z^{\rm sc}$ at the frequencies $\oI$ and $\oII$ in the case of a single gyrotropic rod. Each of the presented fields has a pronounced helical structure and differs significantly from the field scattered by an isotropic rod at the resonant frequency $\o = \o_{\rm r}$ [see Fig.~\ref{fig3}(c)]. It is seen that an isotropic rod scatters the incident wave rather weakly in the direction perpendicular to the wave vector of the incident wave, whereas the scattering pattern of a gyrotropic rod is much more uniform. Far from the rod, this pattern resembles the spatial structure of the $m=0$ azimuthal harmonic of the field scattered by an isotropic rod [see Fig.~\ref{fig3}(d)].

\begin{figure}[h]
\includegraphics[height=70mm]{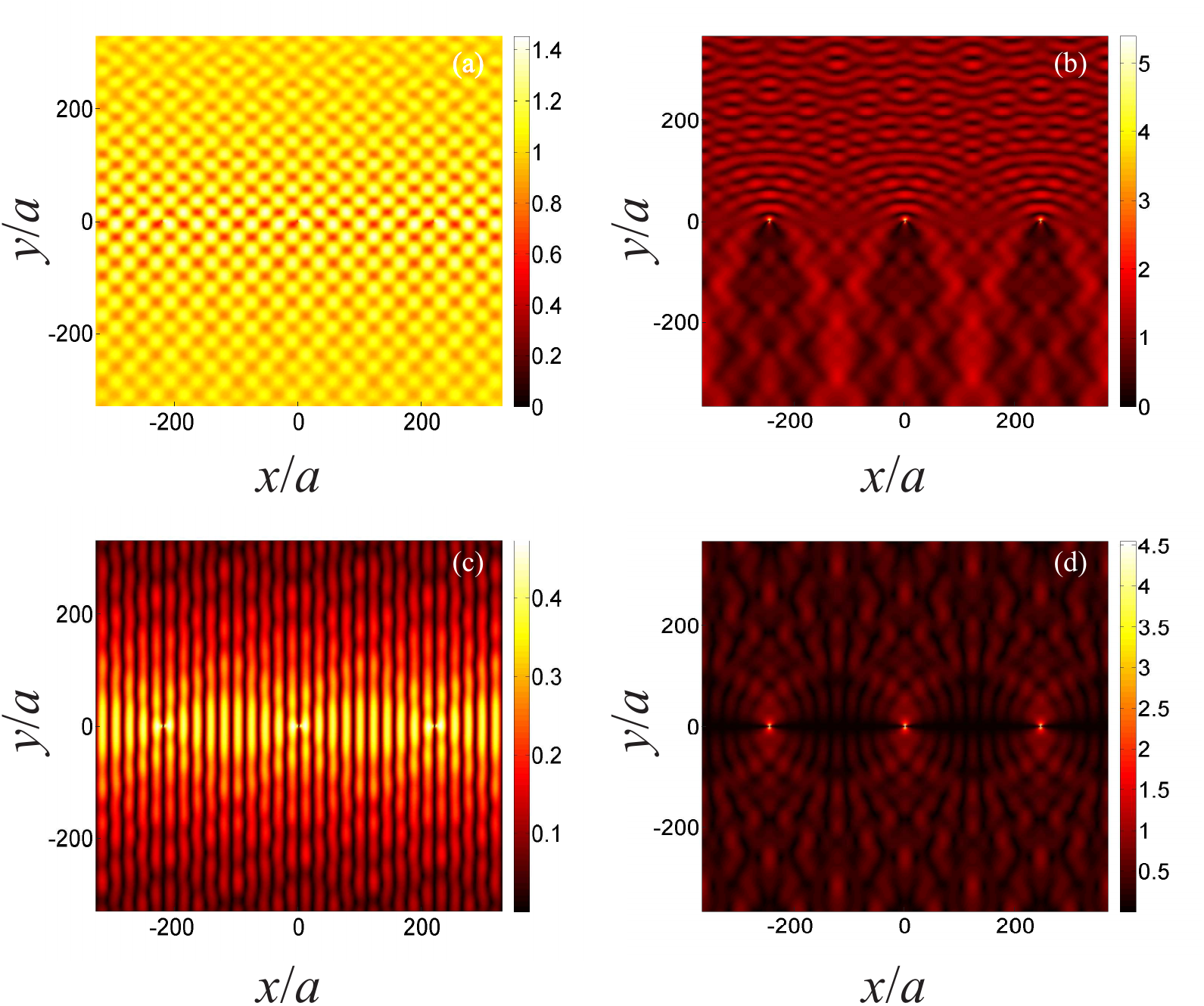}
\caption{\label{fig4}(color online) Absolute value of the total magnetic field $H_z$ for the arrays of (a) gyrotropic and (b) isotropic rods at $\o = \oII$ and $\o = \orr$, respectively, if $kL/2\pi = 4{.}999$, $L/a = 219{.}565$, and $\theta = \pi/2$. The absolute value of the scattered magnetic field $H_z^{\rm sc}$ for the arrays of (c) gyrotropic and (d) isotropic rods at the respective frequencies indicated above. The same values of the parameters $\op/\oH$, $\op a/c$, and $\tilde{\varepsilon}$ as in Fig.~\ref{fig2}.}
\end{figure}

The above-mentioned features directly affect the multiple scattering by an array of cylindrical rods. For example, the absolute value of the total field $H_z$ shown in Fig.~\ref{fig4}(a) for the case of scattering of a plane wave at the frequency $\o = \oII \simeq \o_5$ by the array of gyrotropic rods has a periodic structure in the form of a chessboard. The absolute value of the total field in the case of scattering by the array of isotropic rods at $\o = \orr \simeq \o_5$ does not possess such a spatial structure [see Fig.~\ref{fig4}(b)]. To better clarify the difference between the both cases, it is instructive to compare the corresponding scattered fields, which are shown in Fig.~\ref{fig4}(c) and \ref{fig4}(d). The absolute value of the field scattered by the array of gyrotropic rods is periodic along the $x$-axis and slowly varies in the direction of the $y$-axis. This is related to an almost axisymmetric far-field scattering pattern of each of the gyrotropic rods forming such an array. If  the field scattered by an array of isotropic rods were dominated by the $m=0$ azimuthal harmonic, then this field would by very similar to that shown in Fig.~\ref{fig4}(c). However, such a spatial structure cannot by observed for the array consisting of isotropic rods, because their individual scattered fields are of dipole nature and form another pattern presented in Fig.~\ref{fig4}(d).

\begin{figure}[h]
\includegraphics[height=33mm]{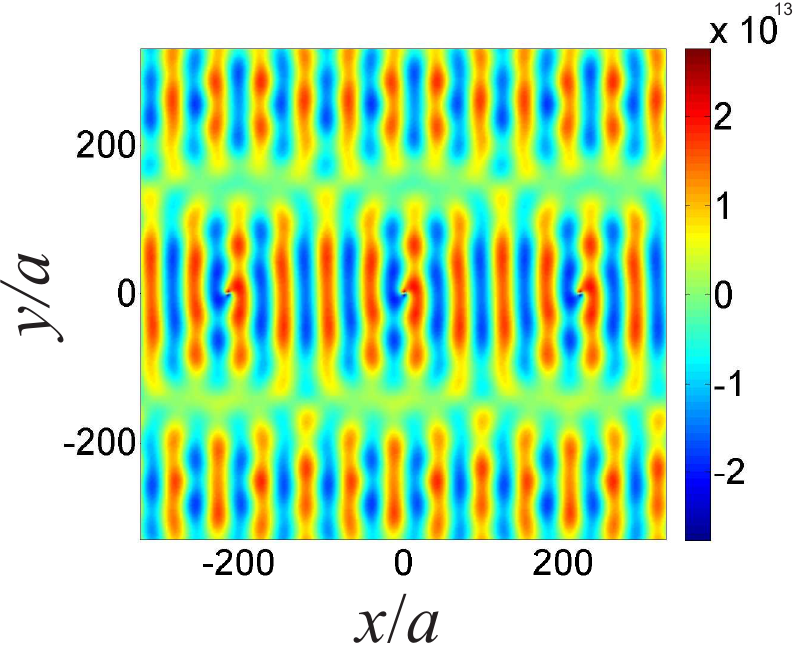}
\caption{\label{fig5}(color online) Snapshots of the total magnetic field $H_z$ for the array of gyrotropic rods at $\o = 0{.}998\oI$ ($\oI = 4{.}0645 \oH$), $kL/2\pi = 4{.}0296$, $L/a = 220{.}2705$, and $\theta = \pi/2$. The same values of the parameters $\op/\oH$, $\op a/c$, and $\tilde{\varepsilon}$ as in Fig.~\ref{fig2}.}
\end{figure}

Under certain conditions, one of the coefficients $\hat{D}_{-1}$ and $\hat{D}_{1}$ can increase extremely greatly, so that the total field is predominantly determined by the scattered field. This situation is depicted in Fig.~\ref{fig5} and takes place when both the real and imaginary parts of the denominator in Eq.~(\ref{eq8}) simultaneously tend to zero. Such a giant enhancement of the scattered field is observed in the case of a small mismatch between any of the resonant frequencies $\omega_{\nu}$ of the array and one of the single-rod resonant frequencies ($\oI$ or $\oII$), provided that $\omega$ is close to the corresponding frequency $\omega_{\nu}$. For the plot of Fig.~\ref{fig5}, $\omega \simeq \oI \simeq \omega_{4}$ and $|\hat{D}_{-1}|\gg |\hat{D}_{1}|$.

\begin{figure}[h]
\includegraphics[height=33mm]{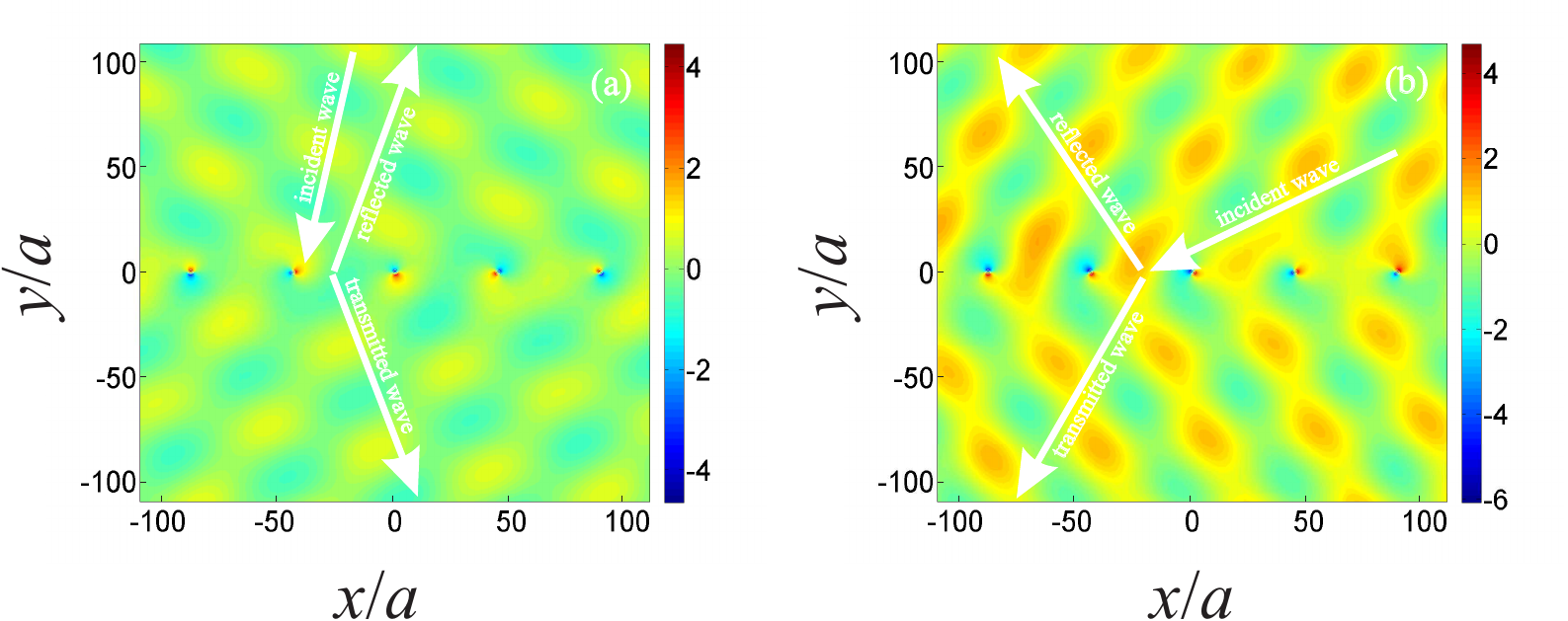}
\caption{\label{fig6}(color online) Snapshots of the scattered field $H_z^{\rm sc}$ in the cases of oblique incidence for (a) $\theta = 3 \pi/7$ and (b) $\theta = \pi/7$, when $k L / 2\pi = 1{.}0002$. The values of other parameters are the same as in Fig.~\ref{fig2}.}
\end{figure}

In the case of oblique incidence where $|\theta|\neq \pi/2$, the scattering demonstrates other interesting features, in addition to those discussed above. Figures~\ref{fig6}(a) and \ref{fig6}(b) show the field distributions for the incidence angles $\theta = 3 \pi/7$ and $\theta = \pi/7$, respectively, if $\o = \oI \simeq \o_1$. The arrows in the figures shows the energy flow directions in the incident wave and in the reflected and transmitted far-zone fields. It is found that for $\theta = 3 \pi /7$ [see Fig.~\ref{fig6}(a)], the reflected and transmitted rays demonstrate the behavior which is essentially different from the more habitual situation depicted in Fig.~\ref{fig6}(b) for $\theta = \pi/7$. Note that Fig.~\ref{fig6}(a) resembles the pattern observed during the resonance excitation of a negative-order spatial harmonic in the case of wave scattering by isotropic periodic structures~\cite{Vlasov2000,Du2011}.

In conclusion, we note that the results obtained can be useful in creating media with 2D distributed feedback~\cite{Baryshev2007}, developing promising photolithography methods~\cite{Koenderink2007}, operating multi-tube helicon plasma sources~\cite{Chen1997}, and understanding the features of wave scattering by periodic magnetic-field-aligned plasma density irregularities in the ionosphere~\cite{Sonwalkar2004,Leyser2009}.

This work was supported by the Government of the Russian Federation (Project No. 11.G34.31.0048),
the RFBR (Project No. 12--02--00747-a), the Russian Federal Program ``Kadry''
(Contracts No.~P313 and No.~02.740.11.0565), the Dynasty Foundation, and
the CNRS (PICS Project No.~4960). A.\,V.\,K. also acknowledges partial support from the
Greek Ministry of Education under the project THALIS (RF--EIGEN--SDR).

\end{document}